\documentclass[12pt]{article}
\usepackage{epsfig,amssymb,latexsym,amsmath}
\usepackage{amssymb}  
\usepackage{graphics}
\usepackage{color}
\usepackage{hyperref}\usepackage{bookmark}

\newcommand{\mb}{\mathbf}
\newcommand{\Sect}[1]{Sect.~\ref{#1}}
\newcommand{\Sects}[2]{Sects.~\ref{#1},~\ref{#2}}
\newcommand{\eq}[1]{(\ref{#1})}
\newcommand{\Eq}{Eq.~\eq}
\newcommand{\be}{\begin{equation}}
\newcommand{\ee}{\end{equation}}

\title{Dynamics-generating semigroups and phenomenology of decoherence}
\author{Michael B. Mensky\\
{\small P.N.Lebedev Physical Institute,} 
{\small 53 Leninsky prosp., 119991 Moscow, Russia}}

\date{09.12.2013}
 
\begin{document}

\maketitle

\begin{abstract}
The earlier proposed Dynamics-Generating Approach (DGA) is reviewed and extended. Starting from an arbitrarily chosen group or semigroup which have structure similar to the structure of Galilei group, DGA allows one to construct phenomenological description of dynamics of the corresponding ``elementary quantum object'' (a particle or non-local object of special type). A class of Galilei-type semigroups, with semigroup of trajectories (parametrized paths) instead of translations, allows one to derive Feynman path integrals in the framework of DGA. The measure of path integrating (exponential of the classic action) is not postulated but derived from the structure of projective semigroup representations. The generalization of DGA suggested in the present paper allows one to derive dynamics of open quantum systems. Specifically, phenomenological description of decoherence and dissipation of a non-relativistic particle is derived from Galilei semigroup. 
\end{abstract}




\section{Introduction}
\label{sec:IntroductionDynamicsGeneratingApproach}

It is well-known that dynamics of a physical system is restricted if this system possesses some symmetry \cite{1,2,3}. The formalism which is appropriate for describing symmetries, is group theory, and behavior of symmetric quantum systems is presented with the help of group representations. 

It is less known that \emph{dynamics} of elementary quantum objects (such as elementary particles but also some types of non-local objects) are not only restricted by symmetry of these objects but \emph{can be derived from the group-theoretical considerations}. Dynamics may be derived from the given group or semigroup \cite{MMbook1976,TMF1983,TMF2012} in the approach that can be called \emph{dynamics-generating approach}, or DGA. 

The resulting dynamics depends on the choice of the group or semigroup that is a starting point of the procedure of DGA. The simplest choice is Galilei group, but in the general case the \emph{dynamics-generating semigroup} (DGS) should have structure similar to the structure of Galilei group, in particularly has to include transformation similar to proper Galilei transformations or their generalizations. One may say that DGS should be a \emph{Galiei-type group or semigroup}. 

Taking \emph{Galilei group} as DGS leads to theory of non-relativistic quantum particles \cite{MMbook1976}. Theory of relativistic particles follows \cite{TMF2012} from the so-called \emph{Aghassi-Roman-Santilly group} \cite{5-Galilei-a,5-Galilei-b,5-GalileiReps} which is a relativistic generalization of Galilei group, with the so-called proper time instead of the usual time. 

Path integrals as the mathematical apparatus presenting quantum dynamics follows, in the framework of DGA, from the non-relativistic or relativistic \emph{Galilei-type semigroup}, which includes the semigroup of parametrized paths (trajectories) instead of the usual translation group \cite{TMF1983,TMF2012}. This may result in theory of point particles, as in Feynman path integral theory. However, instead of this, one may obtain in this way a more general type of dynamics, corresponding non-local objects of special type. We call these objects \emph{history-strings}, and they may be used as a sort of fundamental model for presenting confinement of quarks \cite{13}. 

In the present paper we shall expand DGA in such a way that it might describe open quantum systems, so that the resulting dynamics might include decoherence and dissipation. 

It seems unexpected that DGA may lead to phenomenological description of the dynamics of open quantum systems i.e. those which are subject to influence of their environment. However, this proves to be possible. Logics of the construction is based in this case on the fact that 1)~decoherence and dissipation may be presented phenomenologically with the help of \emph{restricted path integrals} (RPI), or \emph{quantum corridors}, and 2)~the dynamics presented by RPI may also be derived in the framework of DGA. 

\paragraph{Dissipation presented by RPI\\}
\label{sec:RPIDissipation}

Decoherence and dissipation of a quantum system is a consequence of interaction of this system with its environment \cite{Zeh73,Zurek82,JoosZeh85deco,Zeh-bk96}. However, decoherence may in various ways be presented phenomenologically (see \cite{MM2003UFN} and references therein). One of the phenomenological approaches is based on the idea that interaction with the environment remains in the latter some information on the state of the system and may therefore be considered as a continuous measurement of the system. One can make use of \emph{restricted path integrals} (RPI), or \emph{quantum corridors}, for describing such a measurement \cite{TMF1983,MM1993Bk}. 

According to RPI approach, integration in Feynman path integral is restricted onto a set of paths (quantum corridor) which presents the readout of the continuous measurement (information remaining in the environment about the state of the system in all time moments). RPI presents therefore dynamics of an open system under action of its environment. However, no explicit model of the environment is used in the RPI approach.  

Approach based on RPI is advantageous in case when a quantum system is given, and the task is in analyzing how the dynamics of this system change under action of various environments. It is much more cumbersome to apply in these situations conventional approach, with considering the system together with its environment and subsequent integrating out the degrees of freedom of the environment. The phenomenological RPI approach essentially simplifies calculations and allows to systematically explore large classes of environments for the given system.

The simplest model of decoherence and dissipation of a point quantum particle that can be obtained in the framework of RPI approach \cite{MM2003UFN}, leads to the phase representation of the path integral (the integral over paths in the phase space) in which the conventional measure of integration (exponential of the action) is substituted by the following functional: 
\begin{align*}
 	{U}^t_{t'}([\mb{p}],[\mb{x}])  &= \exp \left\{ \int^t_{t'} dt\left[
	\frac i\hbar \Big( \mb{p}\,\dot{\mb{x}}-H_0(\mb{p},\mb{x})\Big) \right. \right. \\
	 &  \left. \left. -\kappa \Big( A(\mb{p},\mb{x})-a(t)\Big) ^2-\frac i\hbar \Big(
	\lambda \,a(t)B(\mb{p},\mb{x})+C(\mb{p},\mb{x})\Big) \right] \right\} 
\end{align*}
The first term here (with $H_0=\mb{p}^2/2m$) presents dynamics of the initial quantum system (point particle of mass $m$), the second term is responsible for decoherence arising due to continuous measuring the observable $\hat A$ (with the precision depending on the coefficient $\kappa$), resulting in the value $a(t)$ of this observable at time $t$, and the last term presents dissipation. More complicated regimes of decoherence may be described analogously, differing only by the choice of the corresponding functions in the exponent. 

\paragraph{Semigroup of quantum corridors\\}
\label{sec:SemigroupOfQuantumCorridors}

In the example presented above, the quantum corridor is determined by the function $[a]^t_{t'}$, i.e. by the valued $a(t)$ of the measured observable at each moment of the interval of the measurement. Operation of multiplication is naturally defined for these quantum corridors as $[a]^t_{t''}=[a]^t_{t'}\cdot [a]^{t'}_{t''}$ is taken. Two corridors that have to be multiplied are joined together in such a way that the second corridor starts at the point where the first one ends. With this definition, corridors form a semi-groupoid (since the inverse element is not defined for an arbitrary corridor and not any pair of the corridors may be multiplied). 

However, the semi-groupoid can be converted into a semigroup (in which any pair of elements may be multiplied) if not individual corridors but classes of corridors are considered as elements, each class containing the corridors differing by the general shift of all its points. In the book \cite{MM1993Bk} a group was additionally defined which transforms different corridors (i.e. different results of the continuous measurement) at the given time interval into one another. Finally, a semigroup of corridors may be defined that has a structure resembling the structure of Galilei group. Such Galilei-type semigroup can be used for deriving, in the framework of DGA, the dynamics of decohering or dissipating systems. 

Here we shall obtain the dynamics of this type without postulating path integral but deriving it from the group-theoretical considerations formulated as dynamics-generating approach (DGA). The origin of the DGA process will be the choice of some Galilei-type semigroup (a group in simple case). All elements of dynamics, including the measure of path integrating (exponential of the classical action), will be determined from the chosen Galilei-type dynamics-generating semigroup (DGS).

\paragraph{Plan of the paper\\}
\label{sec:PlanOfThePaper}

General scheme of the Dynamics-generating Approach (DGA) is exposed in \Sect{sec:DynamicsGeneratingGroupsSemigroups}. The simplest example of application DGA to Galilei group and dynamics of non-relativistic particles is given in \Sect{sec:GalileiGroupInDHA}. Replacement of translation group by semigroup of paths and deriving Feynman path integral from Galilei semigroup is outlined in \Sects{sec:PathsGalileiSemigroup}{sec:PropagatorsFromPathGroup}. In \Sect{sec:PhaseSpaceAndDecoherence} it is shown how phenomenology of decoherence and dissipation can be derived in the course of DGA, and the main points of the paper are briefly commented in \Sect{sec:ConcludingRemarks}. 

\section{Dynamics-generating approach (DGA)}
\label{sec:DynamicsGeneratingGroupsSemigroups}

In this section we shall briefly mention the main points of the dynamics-generating approach (DGA) leading from the arbitrarily chosen group or semigroup of Galilei type, $G$, to dynamics of elementary quantum objects (for simplicity, call them particles, although some of them may be non-local). The approach includes constructing two characteristic representations of $G$ as well as intertwining these representations. 

The \emph{elementary representation} $U_{elem}(G)$  describes the particle as a whole (as an elementary object) while the \emph{local representation} $U_{loc}(G)$ describes the same particle in terms of its localization in the appropriate space (for example space-time, but may be another space relating with the space-time in a more complicated way). Intertwining these representations (mapping between carrier spaces of them, $S_1:\;\mathcal{L}_{elem}\rightarrow \mathcal{L}_{loc}$ and  $S_2:\;\mathcal{L}_{loc}\rightarrow \mathcal{L}_{elem}$) allows one to connect the two descriptions (elementary and local) with each other and derive the causal propagator, i.e. the probability amplitude for the particle to transit from one localized state to the other. 

An important (although purely technical) role in this construction is played by induced representations of groups/semigroups that supply universal mathematical instruments for all stages of DGA. Remark that the group/semigroup should have a structure similar to the structure of Galilei group, but different Galilei-type groups/semigroups lead to different kinds of elementary objects and different kinds of dynamics, including even decoherence and dissipation. 

\subsection{Representations of dynamics-generating groups/se\-mi\-groups}
\label{sec:DynamicsFromSemiGroupsOfGalileiType}

The local representation $U_{loc}(G)$ has to be realized in the space $\mathcal{L}_{loc}$ of functions on the configuration space of the particle. The latter may be the space-time for point particles, but can also have more complicated structure for non-local objects (or even for point particles but under external influence). In any case the configuration space may be realized as a quotient space $G/H$ where $H\subset G$ is a subgroup/subsemigroup, and the local representation may be constructed as a representation $U_{loc}(G)=\chi(H)\uparrow G$ of the group or semigroup $G$ induced from the subgroup/subsemigroup $H$ (see below \Sect{sec:InducedRepresentations} about induced representations). 

The elementary representation $U_{elem}(G)$ should be irreducible (for presenting elementary object) and may be realized in any way. However, it is convenient to realize it also as an induced representation $U_{elem}(G)=\kappa(K)\uparrow G$ from an appropriate representation $\kappa$ of an appropriate subgroup/subsemigroup $K\subset G$. 

If the representations $U_{loc}(G)$ and $U_{elem}(G)$ are realized as induced representations, then the operators which intertwine these representations, are obtained in the standard form as it is shown in \Sect{sec:InducedRepresentations}. The operators $S_1$ and $S_2$ are said to intertwine the representations $U_{elem}(G)$ and $U_{loc}(G)$ in one and opposite directions ($S_1\in [U_{elem},U_{loc}]$) and $S_2\in [U_{loc},U_{elem}]$) if the following commutation relations are valid: 
$$
S_1 U_{elem}(g) = U_{loc}(g) S_1, \quad 
S_2 U_{loc}(g) = U_{elem}(g) S_2. 
$$

These operators map the two representations onto each other, conserving the action of the group/semigroup. The product operator $\Pi = S_1 S_2$ intertwines the local representation $U_{loc}(G)$ with itself. The kernel $\Pi(x'',x')$ of the operator $\Pi$ is then a two-point function with the arguments in the localization space (space-time in the simplest case). This function is interpreted as a probability amplitude for transition from the point $x'$ in the localization space to another point $x''$ of this space. Causal propagator is obtained then by imposing additional condition that the propagation occurs from the past to the future. 

This physical interpretation of the operator $\Pi = S_1 S_2$ is elaborated in the book \cite{MMbook1976} for point non-relativistic and relativistic particles (when localization space is the space-time) and applied in \cite{TMF1983,TMF2012} for deriving non-relativistic and relativistic path integrals (when the localization space is a space of trajectories).  

Different choices of the dynamics-generating groups/semigroups lead to different dynamics: 
	\begin{itemize}
		\item Galilei group:    Non-relativistic particles 
		\item Aghassi-Roman-Santilli group:    Relativistic particles
		\item Galilei-type semigroup (with paths instead of tranlation vectors):    Feynman path integrals
	\end{itemize}
Different choices of configuration spaces describe different physical influences of environment on the particles: 
	\begin{itemize}
		\item Space-time:   Gauge fields (including confinement of color) 
		\item Phase space:    Decoherence and dissipation
	\end{itemize}
Some of these constructions will be briefly considered in the next sections. 

\subsection{Induced representations}
\label{sec:InducedRepresentations}

We shall use so-called induced representations of groups/semigroups as a convenient mathematical instrument. Induced representations generalize the well-known regular representation (acting by left shifts in the space of number-valued functions on the group). The representation $\kappa(K)\uparrow G$ of the group or semigroup $G$ induced from the representation $\kappa(K)$ of the subgroup (subsemigroup) $K\subset G$ is also defined by left shifts of functions on $G$, but the space of these functions (a carrier space of $\kappa(K)\uparrow G$) is defined in a more complicated way. 

The carrier space of the representation $U_\kappa(G)=\kappa(K)\uparrow G$ is defined as a space of functions on $G$ with values in the carrier space $\mathcal{L}_\kappa$ of the representation $\kappa(K)$ and satisfying the so-called \emph{structure condition} 
\be
\varphi(gk) = \kappa(k^{-1})\varphi(g)
\ee
(for any $g\in G$, $k\in K$. The operators of the induced representation $U_\kappa$ act in this space by left shifts:
\be
U_\kappa(g)\varphi(g')=\varphi(g^{-1}g')
\ee
Operators $U(g)$ are unitary in respect to the scalar product
\be
(\varphi, \varphi')=\int_{G/K}\langle\varphi(x_G), \varphi'(x_G)\rangle dx
\ee

\paragraph{Intertwining $S\in[\kappa(K)\uparrow G, \chi(H)\uparrow G]$:\\}
\label{sec:Intertwining}

Operator which intertwines two induced representations of a group/semigroup $G$ have form 
\be
	\label{IntertwineGroup}
	\left(S\varphi\right)(g) = \int_{G} s(g')\varphi(gg')dg'
\ee
	were operator-valued function $s(g):\; \mathcal{L}_\kappa \rightarrow \mathcal{L}_\chi$ satisfies the following two-sided structure condition:
$$\label{IntertwineGroupStructureCondition}
	s(hgk) = \chi(h)s(g)\kappa(k), \quad 
	\forall\, k\in K, \, h\in H
$$
Because of this condition, the integrand is in fact constant on the cosets $gK$. Therefore, integration may be performed over the quotient space $G/K$ rather than over the whole group/semigroup: 
\be
	S\varphi(g) = \int_{G/K} s(x_G)\varphi(gx_G)dx 
\ee

\section{Galilei group and non-relativistic particles}
\label{sec:GalileiGroupInDHA}

The simplest case of applying DGA is the derivation of dynamics of non-relativistic particles from the Galilei group \cite{MMbook1976}. We shall briefly expose here the scheme of this construction. The more complicated applications in the following sections will follow the same general scheme. 

\subsection{Galilei group}
\label{sec:GalileiGroup}

\paragraph{Elements of Galilei group}
\label{sec:ElementsOfGalileiGroup}
may be presented as products  ($g = a_T r\mb{v}_L$) of translations, rotations and proper Galilei transformations. Each of these are defined by their action on the space-time points $x=\{x^0, \mb{x}\}$ as follows (notations are evident):
\begin{align*}
	a_T x &= x+a =\{x^0+a^0, \mb{x}+\mb{a}\} \\
	r x &= \{x^0, r\mb{x}\}, \quad 
	\mb{v}_G x = \{x^0, \mb{x} + x^0\mb{v}\}
\end{align*}

\paragraph{Multiplications in Galilei group} 
\label{sec:MultiplicationsAndCommutationsInGalileiGroup}
are completely determined by the following relations: 
\begin{align*}
	a_T a'_T = (a+a')_T, \quad &\mb{v}_G \mb{v'}_G = (\mb{v}+\mb{v'})_G \\
	r \mb{v}_G r^{-1} = (r\mb{v})_G, \quad 
	&\mb{v}_G a_T \mb{v}_G^{-1} = (\mb{v}_G a)_T 
	= \{a^0, \mb{a} + a^0\mb{v}\}_T 
\end{align*}

\subsection{Projective representations as an origin of dynamics}
\label{sec:ProjectiveRepresentations}
\paragraph{Projective representations}
\label{sec:Multiplicators}
of the Galilei group are defined as 
$$
U(g)U(g') = \lambda(g, g')U(gg')
$$
where $\lambda(g, g')$ are complex numbers called \emph{multiplicators of the given representation}. The system of multiplicators for a projective representation of Galilei group is determined as 
$$
\lambda\big(a_T r\mb{v}_G , a'_T r' \mb{v'}_G \big) 
= \lambda\big(\mb{v}_G , a'_T  \big) 
= \exp\left[im\left(\mb{v}\mb{a}' + \frac{1}{2}\mb{v}^2 a'^0\right)\right] .
$$
It depends on a single parameter $m$ (which will play the role of mass in the dynamics resulting from DGA). 

\paragraph{Propagator following from Galilei group}
due to DGA (i.e. as $\Pi=S_1S_2$ with the operators $S_i$ intertwining elementary and local representations) is equal (up to a number factor) \cite{MMbook1976} to 
\begin{align*}
	\left(\Pi\psi\right)(x)&=\int d^4x'\,\Pi(x-x')\psi(x') , \\
	\Pi(x-x')&=\int d\mb{v}\,
	\exp\left\{i\left[m\mb{v}(\mb{x}-\mb{x'})-\frac 12 m\mb{v}^2(x^0-x'^0)\right]\right\} .
\end{align*}
\textbf{Causal propagator} is obtained then if one requires that the transition is performed from the past to the future: 
$$
	\Pi^c(x-x')=\theta(x^0-x'^0)\Pi(x-x') .
$$

\section{Paths and Galilei semigroup}
\label{sec:PathsGalileiSemigroup}

\emph{Galilei semigroup} is obtained if translation subgroup of Galilei group is replaced by the semigroup of parametrized paths, or trajectories \cite{13,MM1993Bk,7,MM1979}. Dynamics resulting in the framework of DGA \cite{TMF1983} is what is known as Feynman path integrals. It will be considered in \Sect{sec:PropagatorsFromPathGroup}. 

\subsection{Paths instead of translations}
\label{sec:PathsInsteadOfTranslations}

The semigroup of trajectories (parametrized paths) is defined as a generalization of the translation group. This may be done in the following steps. 
\begin{figure}[h]
	\centering
		\includegraphics[width=0.6\textwidth]{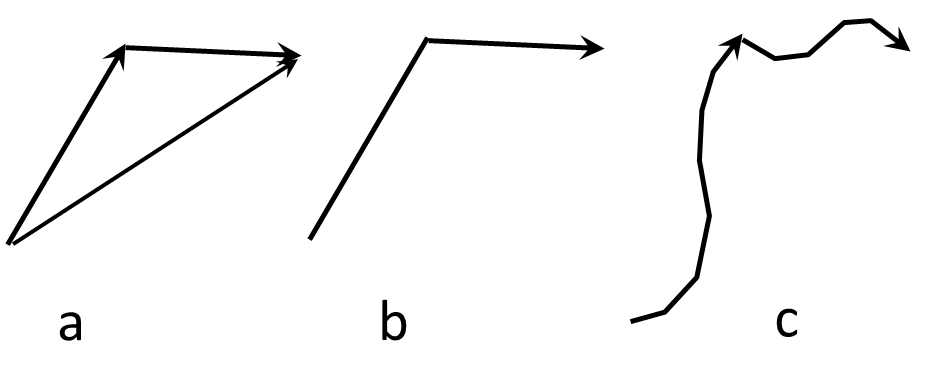}
	\caption{Paths as generalization of translations}
	\label{fig:PathGrDef}
\end{figure}

\begin{itemize}
	\item Translations as elements of the translation group are defined as vectors (in the corresponding vector space), see Fig.~\ref{fig:PathGrDef}~a. Product of two translations is determined by the sum of the corresponding two vectors. 

	\item Let us take a generalized translation along some vector and another generalized translation, along another vector. Let us define the product of these generalized translations not as a sum of the two vectors, but as a prolongation of the first vector by the second vector, see Fig.~\ref{fig:PathGrDef}~b. 

	\item Consider a set of continuous curves in some linear space. Present each curve as a consequence of infinitesimal vectors prolonging one another. Define product of two continuous curves as prolongation of one of them by the other (provided that the beginning point of the second curve coincides with the end point of the first). This gives what can be called \emph{the groupoid of curves} (groupoid because not any pair of continuous curves may be multiplied). It is evident that each continuous curve may be considered then as a generalized translation along this curve. 
	
	\item Going over from the individual curves to the classes of curves, we can define \emph{the semigroup of parametrized paths (trajectories)}. Two curves are equivalent if all points of one of them are obtained from the corresponding points of the other by shifting along the same vector. Product of two paths are defined then as prolongation the curve belonging to one of them by the curve belonging to the other, see Fig.~\ref{fig:PathGrDef}~c. 
\end{itemize}

If the paths are parametrized by the time parameter, we shall denote any curve $\mb{x}(t)$ belonging to this class as $[\mb{x}]_{t'}^{t''}$ where $\mb{x}(t')$ and $\mb{x}(t'')$ are its initial and final points. The path is defined as a class of equivalent curves, two curves  being equivalent if all points of one of them differ from the points of the other curve by the same vector. We have thus the \emph{semigroup of paths} (the term ``path'' denoting a class of equivalent curves). Because of the equivalence, the path, i.e. the class of equivalent curves, is defined by velocities $\dot{\mb{x}}(t)$. We shall denote such a class as $[\mb{x}]_{t'}^{t''} = [\dot{\mb{x}}]_{t'}^{t''}$ or even $[\mb{x}]_{t'}^{t''} = [\mb{u}]_{t''-t'}$ where $\mb{u}(t)=\dot{\mb{x}}(t-t')$. 

\subsection{Structure of Galilei semigroup}
\label{sec:StructureOfAGalileiTypeSemigroup}

Galilei semigroup is defined if the translations of Galilei group are replaced by trajectories (or parametrized paths) $[\mb{x}]_{t'}^{t''} = [\mb{u}]_{t''-t'}$. Correspondingly, the elementary quantum object that will be described in the framework of DGA should have the space of trajectories as its localization space. 

Generalization of a proper Galilei transformation is determined as family of velocities, one velocity for each time: $[\mb{v}]=\{\mb{v}(t)|-\infty <t<\infty\}$ with products defined trivially: 
\be\label{ProperGalilei}
[\mb{v}][\mb{v}'] = [\mb{v}+\mb{v}'] = \{\mb{v}(t)+\mb{v}'(t)|-\infty <t<\infty\}.
\ee

Besides the mentioned elements, Galilei semigroup contains rotations $r\in R$, with the evident relations: 
\be\label{rotations}
r[\mb{v}]r^{-1} = [r\mb{v}] , \quad 
r[\mb{u}]_\tau r^{-1} = [r\mb{u}]_\tau, 
\ee
The Galilei semigroup as a whole contains these elements as well as their products, $g=[\mb{u}]_\tau r[\mb{v}]$. 
\begin{figure}[h]
	\centering
	\includegraphics[width=0.80\textwidth]{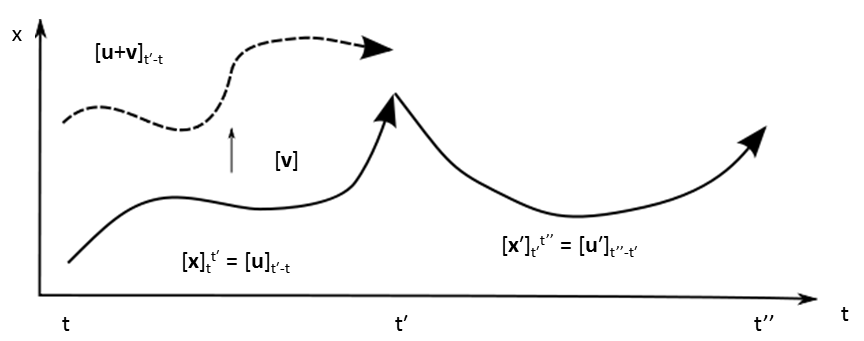}
	\caption{Structure of Galilei semigroup}
	\label{fig:GalileySemigroupStructure}
\end{figure}
Interrelations of the characteristic elements of Galilei semigroup are illustrated at Fig.~\ref{fig:GalileySemigroupStructure}. As has been said previously, product of paths (trajectories) $[\mb{x}]_{t'}^{t''}[\mb{x}]_{t}^{t'}$ is defined as prolonging (extending) them by each other. The dashed trajectory illustrates action of a proper Galilei transformation $[\mb{v}]$ on trajectories. This action converts the paths on the given time interval into each other, which is expressed by the following commutation relation: 
\be\label{GalileiCommute}
[\mb{v}][\mb{u}]_\tau [\mb{v}]^{-1} = [\mb{u}+\mb{v}]_\tau
\ee

The Galilei semigroup (just as Galilei group) has (non-trivial) projective representations, with the systems of multiplicators of the form
\be\label{GalileiSemigroupMultiplicator}
\Big([\mb{u}]_\tau r[\mb{v}] , [\mb{u}']_\tau r'[\mb{v}']\Big) 
= \exp\left[im \int_0^\tau d\sigma \left(\mb{uv}' + \frac{1}{2}\mb{v}'^2\right)\right] .
\ee

Instead of projective representations of Galilei semigroup one can consider only usual vector representation but for central extension of the semigroup (we shall denote it by the same letter $G$). Elements of the extended semigroup are defined as $g=\lambda [\mb{u}]_\tau r[\mb{v}]$ where $\lambda$ is an arbitrary complex number (therefore, an element from the center of the semigroup $G$). For elements of the extended semigroup the same relations are valid but with \Eq{GalileiCommute} replaced by the following: 
\begin{align*}\label{GeneralizedGalileiCommute}
	[\mb{v}][\mb{u}]_\tau[\mb{v}]^{-1}	&=	\lambda \cdot [\mb{u}+\mb{v}]_\tau, \quad
\lambda 
= \exp\left[im \int_0^\tau d\sigma \left(\mb{u}\mb{v} + \frac{1}{2}\mb{v}^2\right)\right]  
\end{align*}

\section{Propagators in DGA with paths instead of translations}
\label{sec:PropagatorsFromPathGroup}

If we start the procedure of DGA from the Galilei semigroup, we obtain finally \cite{TMF1983} the dynamics presented by Feynman path integrals. The concept of path integrals, including the measure of integrating (the famous exponential of the classical action) is not postulated in this case but derived from the group-theoretical considerations. 

In the context of DGA, Galilei semigroup leads to the space of trajectories as a localization space. In principle, this allows one to obtain the dynamics of non-local objects.\footnote{Such non-local objects can naturally present confinement of quarks \cite{13}.} If our goal is theory of point particles, then we make use of the propagator in the space of trajectories (\Sect{sec:CausalPropagatorInTheSpaceOfTrajectories}) and go over to the propagator in the space-time (as is shown in \Sect{sec:PropagatorInSpaceTime}).

\subsection{Causal propagator in the space of trajectories}
\label{sec:CausalPropagatorInTheSpaceOfTrajectories}

In \Sect{sec:DynamicsFromSemiGroupsOfGalileiType} the general procedure (DGA) is exposed for constructing causal propagator of the elementary quantum object with the help of intertwining local and elementary representations of the dynamics-generating group/semigroup. This procedure consists in i)~constructing intertwining operators $S_1\in [U_{elem},U_{loc}]$, $S_2\in [U_{loc},U_{elem}]$, and $\Pi = S_1 S_2$, ii)~finding the kernel $\Pi(x'',x')$ of the latter and iii)~imposing the causality condition for obtaining the causal propagator $\Pi^c(x'',x')$. 

In case of Galilei semigroup, with paths instead of space-time translations, the role of localization space is played by trajectories (parametrized paths) \cite{TMF1983,TMF2012}. The resulting form of the causal propagator, up to a number factor, is then \cite{TMF1983}
\be\label{CausalPropGalSemigr}
	(\Pi^c\psi)[\mb{x}]_0^t 
		= \theta(t-t')\int_0^t d{t'} \, 
		\int d[\mb{p}]_{t'}^t \cdot 
		U_{t'}^{t}([\mb{p}],[\mb{x}])\cdot \psi[\mb{x}]_0^{t'}
\ee
where it is denoted 
\be\label{KernelSimple}
		U_{t'}^{t}([\mb{p}],[\mb{x}])
		= \exp\left[\frac{i}{\hbar} 
		\int_{t'}^t d{t} \left(\mb{p}\mb{\dot{x}} - H_0(\mb{p})\right) \right]
\ee
with 
$$
		H_0(\mb{p}) 
		= \frac{1}{2m}\mb{p}^2
$$

This form of the propagator is formally derived from intertwining the local and elementary representations. We see that the Feynman measure in the space of trajectories is not postulated (as an exponential of the classical action) but derived from the group-theoretical considerations, namely from the system of multiplicators \eq{GalileiSemigroupMultiplicator} of Galilei semigroup. 

\subsection{Propagator in space-time}
\label{sec:PropagatorInSpaceTime}
In \Sect{sec:CausalPropagatorInTheSpaceOfTrajectories} we considered the space of trajectories (parametrized paths) as a localization space of our elementary quantum object. Therefore, this object may be in principle non-local. Such non-local objects (called `history-strings') were considered in \cite{TMF2012,13}. However, dynamics of a point particle can also be derived from the path-integral formalism obtained in \Sect{sec:CausalPropagatorInTheSpaceOfTrajectories}. 

For this aim, we have to define the probability amplitude to be in a definite space-time point as a sum of the amplitudes to arrive to this point along various paths (trajectories). 

The action of paths on the space-time may be naturally defined as 
$$
(\mb{x}'',t'') = [\mb{x}]^{t''}_{t'}(\mb{x}',t')
$$
where $(\mb{x}',t')$ and $(\mb{x}'',t'')$ are correspondingly the initial and final space-time points of the path $[\mb{x}]^{t''}_{t'}$. Let us say in this case that the point $(\mb{x}'',t'')$ is arrived from $(\mb{x}',t')$ along the path $[\mb{x}]^{t''}_{t'}$. 

Let us suppose\footnote{This assumption may be generalized, but this is not essential for us now.} that 1)~at the time moment $t=0$ the point particle is localized in a single point $(\mb{x}_0,0)$ and 2)~it can arrive to an arbitrary space-time points $(\mb{x}',t')$ along paths $[\mb{x}]^{t'}_{0}$ according to the formula
$$
(\mb{x}',t') = [\mb{x}]^{t'}_{0}(\mb{x_0},0). 
$$
Let the wave function of the particle in the space of trajectories is $\psi[\mb{x}]^{t''}_{t'}$. This means that the probability amplitude for the particle to move along the path $[\mb{x}]^{t''}_{t'}$ is equal to $\psi[\mb{x}]^{t''}_{t'}$. Then the amplitude for this particle to arrive to the point $(\mb{x}',t')$ along the path $[\mb{x}]^{t'}_{0}$ is equal to $\psi[\mb{x}]^{t'}_{0}$. Therefore, the amplitude to be at this point (i.e. to arrive to this point along any path leading to it) is equal to 
\be\label{FromPathsToPoints}
\Psi(\mb{x}',t') = \int_{\mb{x}''}^{\mb{x}'} d[\mb{x}]^{t'}_{0}  
\alpha[\mb{x}]^{t'}_0 \psi[\mb{x}]^{t'}_{0}, \quad \text{where} \quad
(\mb{x}',t') = [\mb{x}]^{t'}_{0}(\mb{x}_0,0)
\ee
with some weight function $\alpha[\mb{x}]^{t'}_0$. This is nothing else than a formula for transition from the wave function (of a point particle) given as a function on the space of trajectories, to the wave function on the space-time. 

The weight function $\alpha[\mb{x}]^{t'}_0$ is not arbitrary. In order for the formula \eq{FromPathsToPoints} to be in accord with the multiplicative structure of the paths, the following condition should be valid,  
$$
\alpha[\mb{x}]_{t'}^{t}\cdot \alpha[\mb{x}]^{t'}_{t''} 
		= \alpha[\mb{x}]^{t}_{t''}, 
		\quad \text{where} \quad
		[\mb{x}]_{t'}^{t}\cdot [\mb{x}]^{t'}_{t''} 
		= [\mb{x}]^{t}_{t''} .
$$
In other words, the function $\alpha$ has to be a representation of the semigroup of trajectories. It is easy to show \cite{TMF1983} that the most general form of the weight function with these properties is 
$$
\alpha\{\mathbf{x}\}^{t'}_{t''} 
= T \exp\left\{i\int^{t'}_{t''}  dt\, \left[V(\mathbf{x}(t),t) 
+ \mathbf{A}(\mathbf{x}(t),t)\dot{\mathbf{x}}\right]\right\} .
$$
where ``$T\exp$'' denotes time-ordered exponential, $V$ a potential and $\mathbf{A}$ a gauge field. 

Equation (\ref{CausalPropGalSemigr}) described causal propagation in terms of path-dependent wave function. Going over, for a point particle, to the point-dependent wave function (\ref{FromPathsToPoints}), we have the point-dependent form of the causal propagator \cite{TMF1983}
\begin{align*}
	(\Pi^c\Psi)(\mb{x},t)
	&= \theta(t-t')\int_{0}^{t} dt' \int d\mb{\tilde x} 
	\int_{\mb{\tilde x}}^{\mb{x}} d[\mb{x}]^{t}_{t'}\cdot \alpha[\mb{x}]_{t'}^{t} \int d[\mb{p}]_{t'}^{t}\\
	& \times \exp\left\{\int_{t'}^{t} dt\left[\frac{i}{\hbar}\left(\mb{p}\mb{\dot{x}} - H_0(\mb{p}\right)\right] \right\}\cdot \Psi(\mb{\tilde x},t') .
\end{align*}
It coincides with the usual path-integral form of non-relativistic causal propagator.

\section{Phase space and decoherence in DGA}
\label{sec:PhaseSpaceAndDecoherence}

Dynamics-generating approach (DGA) is aimed at the derivation of dynamics of elementary quantum objects, for example elementary particles. It seems at first glance that dynamics of complicated physical systems hardly can be derived in the same way. Even more strange to believe that dynamics of open quantum systems can be obtained in the framework of DGA. 

The system is called open if its interaction with the environment affects its dynamics. As a result of this influence, the open quantum system exhibits the phenomenon called decoherence. In the course of gradual decoherence, the quantum system partially loses the quantum character of its evolution, behaves more like a classical system. Dissipation is an extreme form of decoherence. 

It is unexpected that dynamics of open quantum systems, including phenomena of decoherence and dissipation, also can be derived in the framework of DGA. From the mathematical point of view, the price paid for this, is necessity to consider the (generalized) phase space as a localization space for our ``elementary quantum object'' (although this term becomes questionable in this situation).

In fact, this is not so strange because open quantum systems can be described phenomenologically, so that the environment is not explicitly included in the description (but its affect is taken into account implicitly). Here we shall remind the phenomenological description of decoherence by restricted path integrals (RPI) \cite{TMF1983,MM1993Bk} and then derive the same dynamics in the framework of DGA.  

\subsection{Decoherence in RPI approach}
\label{sec:PhenomenologyOfDecoherence}

Decoherence as an effect of the environment can be taken into account directly in models similar to the widely known Caldeira-Leggett model \cite{CaldeiraLegg83}, or with the help of the Feynman-Vernon influence functional included in the path integral  \cite{FeynVernon63}. Instead of this, one may use the phenomenological Lindblad equation \cite{Lindblad76}. We shall prefer to make use of the RPI approach (see \cite{TMF1983,MM1993Bk} on continuous quantum measurements in terms of RPI and \cite{MM2003UFN} on its application to decoherence and dissipation). 

The role of environment may be considered to be a sort of continuous measurement, and dynamics of continuously measured quantum system may be accounted by restricting its path integral onto the corresponding \emph{quantum corridor}, i.e. on the family of paths which corresponds to the result (output) of the measurement. This gives \cite{MM2003UFN} for the propagator of the open (continuously measured) system the expression 
$$
U_\alpha(q'', q') = \int_{q'}^{q''} d[p]d[q]\cdot W_\alpha[p,q]\cdot 
\exp\left\{\frac{i}{\hbar}\int_{t}^{t'}dt\left(p\dot q - H(p,q,t)\right)\right\}
$$
where $\alpha$ denotes the measurement result and the weight functional $W_\alpha[p,q]$ restricts Feynman path integral on the subset of paths corresponding to the measurement result $\alpha$. 

In \cite{MM2003UFN} the author considered the special case of a non-relativistic particle moving through the medium. This may be interpreted as continuous measuring some observable $\hat{A}$ of the particle, with the result of the measurement given by the values $a(t)$ of this observable in various time moments, i.e. by the curve $\alpha = [a]^{t''}_{t'} =\{a(t)|t'<t<t''\}$. This is described by the restricted path integral for the non-relativistic particle, for example in the form \Eq{CausalPropGalSemigr} but with a more complicated kernel $U^{t''}_{t'}$:
\begin{align}\label{KernelDeco}
 	{U}^{t''}_{t'}([\mb{p}],[\mb{x}])  &= \exp \left\{ \int^{t''}_{t'} dt\left[
	\frac i\hbar \Big( \mb{p}\,\dot{\mb{x}}-H_0(\mb{p},\mb{x})\Big) \right. \right. \nonumber\\
	 &  \left. \left. -\kappa \Big( A(\mb{p},\mb{x})-a(t)\Big) ^2-\frac i\hbar \Big(
	\lambda \,a(t)B(\mb{p},\mb{x})+C(\mb{p},\mb{x})\Big) \right] \right\} 
\end{align}

The term $-\kappa(A-a)^2$ here describes the restriction of the path integral that corresponds to the continuous measurement of $A$ with the precision determined by the coefficient $\kappa$. This term is responsible for decoherence. The terms including $B$ and $C$, take into account an additional effect of dissipation that can arise in the process of the measurement.\footnote{The form of the ``decoherence term'' $-\kappa(A-a)^2$ in the exponent of \Eq{KernelDeco} is defined by the very general arguments \cite{MM2000bk}, but additional terms may have a more complicated form.} For example the terms of this type may take into account deceleration by measurement, i.e. dissipation of energy. 

This phenomenological description of decoherence and dissipation has been obtained in terms of restricted path integral. This type of dynamics emerges as a secondary effect rather than is defined on the fundamental level. It is questionable therefore whether it can be derived in the framework of DGA. We shall see in \Sect{sec:PropagatorInPhaseSpace} that this is possible if one introduce the (generalized) phase space as a localization space of our ``elementary quantum object''. 

\subsection{Propagator in phase space}
\label{sec:PropagatorInPhaseSpace}

Propagator in the space of paths is defined by \Eq{CausalPropGalSemigr} with the kernel \eq{KernelSimple}, i.e. in the following form: 
\begin{align*}
	(\Pi^c\psi)[\mb{x}]_0^{{t}} 
			&= \int_0^{t} d{t'} \, 
			\int d[{\mb{p}}]_{t'}^{t} \, \exp \left\{ \int_{t'}^{t} d{t}\left[
		\frac i\hbar \Big( \mb{p}\,\dot{\mb{x}}-H_0(\mb{p})\Big) \right] \right\}
		 \psi[\mb{x}]_0^{t'} 
\end{align*}
We have seen in \Sect{sec:PropagatorInSpaceTime} how the propagator in the space-time can be obtained from this general path-dependent formula. Let us now apply the same path-dependent propagator for deriving \emph{propagator in the (generalized) phase space} and further for derivation phenomenological description of decoherence and dissipation. 

The construction in \Sect{sec:PropagatorInSpaceTime} began by transition from the path-dependent wave function to the space-time-dependent wave function \eq{FromPathsToPoints}. Now we have to go over \emph{from the space of paths to the generalized phase space} rather than to space-time. 

The point of the usual phase space is characterized by the pair ``position and momentum''. We shall denote the point of the generalized phase space as the pair $\left([\mb{p}]_0^{t'};(\mb{x}',t')\right)$ including a space-time point and a path in the momentum space. The wave function in this space has to be obtained with the help of the procedure similar to one applied in \Sect{sec:PropagatorInSpaceTime}. The value of the wave function in the definite point $\left([\mb{p}]_0^{t'};(\mb{x}',t')\right)$ of the generalized phase space is equal to the integral over all paths leading to this point. This results in the wave function of the form 
$$
\Psi\left([\mb{p}]_0^{t'};(\mb{x}',t')\right) 
	= \int_{\mb{x}''}^{\mb{x}'} d[{\mb{x}}]^{t'}_{0}\, 
	\alpha([{\mb{p}}],[{\mb{x}}])^{t'}_{0}\, \psi[{\mb{x}}]^{t'}_{0}  
$$ 
where $\alpha$ is some weight function depending on the paths in the phase space.\footnote{This weight function implicitly accounts for the influence of the environment and therefore determines details of dynamics of the open system.} 

The propagator in the generalized phase space is then derived along the arguments similar to those from \Sect{sec:PropagatorInSpaceTime}: 
\begin{align*}
	(\Pi^c\Psi)\left([\mb{p}]_0^{t};(\mb{x},t)\right) 
			&= \int_0^t d{t'}  \int d[{{\mb{p}}}]_{t'}^t 
			\int d[{\mb{x}}]^{t}_{0} \\
			&\times \alpha^{t}_{0}([{\mb{p}}],[{\mb{x}}]) \cdot \exp \left\{ \int_{t'}^t dt\left[
		\frac i\hbar \Big( \mb{p}\,\dot{\mb{x}}-H_0(\mb{p})\Big) \right] \right\}\,
			\Psi\left([\mb{p}]_0^{t'};(\mb{x}',t')\right) 
\end{align*}
or 
\begin{align*}
	(\Pi^c\Psi)\left([\mb{p}]_0^{t};(\mb{x},t)\right) 
			&= \int_0^t d{t'}  \int d[{{\mb{p}}}]_{t'}^t 
			\int d[{\mb{x}}]^{t}_{0} \cdot 
			{U}_{t'}^t([\mb{p}],[\mb{x}])\,
			\Psi\left([\mb{p}]_0^{t'};(\mb{x}',t')\right) 
\end{align*}
where it is denoted 
\be\label{DecoKernelGeneral}
{U}_{t'}^t([\mb{p}],[\mb{x}]) 
= \alpha^{t}_{0}([{\mb{p}}],[{\mb{x}}]) \cdot \exp \left\{ \int_{t'}^t dt\left[
		\frac i\hbar \Big( \mb{p}\,\dot{\mb{x}}-H_0(\mb{p})\Big) \right] \right\}.
\ee

Depending on the choice of the weight function $\alpha^{t}_{0}([{\mb{p}}],[{\mb{x}}])$, \Eq{DecoKernelGeneral} presents general form of decoherence and dissipation of the non-relativistic particle. In particular, the kernel describing decoherence and dissipation of the form 
\begin{align*}
 	{U}_{t'}^t([\mb{p}],[\mb{x}])  &= \exp \left\{ \int_{t'}^t dt\left[
	\frac i\hbar \Big( \mb{p}\,\dot{\mb{x}}-H_0(\mb{p},\mb{x})\Big) \right. \right. \\
	 &  \left. \left. -\kappa \Big( A(\mb{p},\mb{x})-a(t)\Big) ^2-\frac i\hbar \Big(
	\lambda \,a(t)B(\mb{p},\mb{x})+C(\mb{p},\mb{x})\Big) \right] \right\} 
\end{align*}
(as in \Eq{KernelDeco}) may be obtained if $\alpha$ is chosen as follows:
\begin{align*}
	\alpha\left([\mb{x}]^{t}_{t'},[{\mb{p}}]_{t'}^t  \right) 
	&= \exp \left\{ - \int_{t'}^{t} dt\left[
	\frac i\hbar \Big(+\lambda \,a(t)B(\mb{p},\mb{x})+C(\mb{p},\mb{x})\Big)  +\kappa \Big( A(\mb{p},\mb{x})-a(t)\Big)^2
	 \right] \right\}
	\end{align*}

After integrating in $[\mb{p}]_0^{t}$, we have finally the causal propagator in space-time  
\begin{align*}
		\left(\Pi^c\Psi\right)(\mb{x},t) 
		&= \int_0^t d{t'} \, \int d{\tilde{\mb{x}}}
		\int_{\tilde{\mb{x}}}^{\mb{x}} d[\mb{x}]^{t}_{t'}  
		\int d[{\mb{p}}]_{t'}^t \; 
		\\
		&\times \exp \left\{ \int_{t'}^{t} dt\left[
		\frac i\hbar \Big( \mb{p}\,\dot{\mb{x}}-H_0(\mb{p},\mb{x})
		- C(\mb{p},\mb{x})  - 	\lambda \,a(t)B(\mb{p},\mb{x})
		\Big) \right. \right. \\
		 &  \left. \left. -\kappa \Big( A(\mb{p},\mb{x})-a(t)\Big)^2 \right] \right\}  
		\cdot	\Psi(\tilde{\mb{x}},{t'})
\end{align*}
correctly describing decoherence and dissipation,\footnote{This equation presents the partial effect of decoherence and dissipation corresponding to the measurement result $[a]_{t'}^{t}$. The complete effect may be obtained by going over to the density matrix and integrating over all $[a]_{t'}^{t}$, see \cite{MM2003UFN,MM2000bk}.} i.e. influence of the environment. However, we derived the propagator in the framework of the universal procedure of dynamics-generating approach. The price for inclusion decoherence and dissipation in the scope of this approach is that we came to the propagation in space-time not directly, but through the intermediate form of the propagator in the generalized phase space. 

\section{Concluding remarks}
\label{sec:ConcludingRemarks}

In the preceding we gave a brief review of the Dynamics-Generating Approach (DGA) and demonstrated that it is applicable for derivating dynamics of open quantum systems. 

The goal of DGA is to derive phenomenological description of dynamics of an ``elementary quantum object'' (called ``particle'' although it can be non-local). DGA starts from choosing some group or semigroup $G$ having structure similar to the structure of Galilei group. Then two special representations of this group/semigroup are constructed, $U_{elem}(G)$ presenting the state of the particle as a whole, and $U_{loc}(G)$ describing localization of this state. Intertwining these representations allows one to provide agreement of these two description and derive propagator of the particle in an appropriate localization space (it may be space-time or the space of paths in the space-time). 

It has been earlier shown that, depending on the concrete choice of the dynamics-generating semigroup $G$, the dynamics of a non-relativistic or relativistic local or non-local object can be obtained in the framework of DGA. Feynman path integral is derived in such a way, and the measure of the path integrating (exponential of the classical action) is also derived from $G$ rather than postulated. 

It is additionally shown in the present paper that dynamics of open quantum systems, including phenomena of decoherence and dissipation, can also be derived in such a way. This is achieved if the localization space of the ``particle'' is defined as a generalized phase space and propagation in this space is first derived as an intermediate point of the theory. In such a way the phenomenological description of a non-relativistic particles subjected to decoherence and dissipation is derived. Although this type of dynamics physically emerges as an affect of the environment, in the framework of DGA the environment is not considered explicitly. The decohering particle is considered in this case as a special sort of an elementary quantum object.



\end{document}